\pdfoutput=1
\documentclass[sigconf, nonacm]{acmart}

\AtBeginDocument{%
  }

\setcopyright{acmcopyright}
\copyrightyear{2022}
\acmYear{2022}
\acmDOI{XXXXXXX.XXXXXXX}

\acmConference[MEMSYS '22]{Proceedings of the International Symposium on Memory Systems}{October--December,
  2022}{Virtual Conference}
\acmPrice{15.00}
\acmISBN{978-1-4503-XXXX-X/18/06}

\usepackage{caption}
\usepackage{subcaption}

\definecolor{Black}{HTML}{221E1F}

\definecolor{Red}{HTML}{ED1B23}
\definecolor{Orange}{HTML}{F58137}

\definecolor{Blue}{HTML}{2D2F92}
\definecolor{ProcessBlue}{HTML}{00B0F0}

\definecolor{ForestGreen}{HTML}{009B55}
\definecolor{LimeGreen}{HTML}{8DC73E}

\usepackage{listings}

\usepackage{pgfplots}
\pgfplotsset{compat=1.15}
\usepgfplotslibrary{units}

\definecolor{darkgreen}{rgb}{0,0.43,0}

\lstdefinelanguage{DRAMml}{
    keywords={},
    otherkeywords={
    ->, ->>, -<>, -o, *\\
    },
    keywordstyle=\color{blue}\bfseries,
    keywords=[2]{Hierarchy, Places, Arcs},
    keywordstyle=[2]\color{darkgreen}\bfseries,
    identifierstyle=\color{black},
    sensitive=false,
    comment=[l]{//},
    morecomment=[s]{/*}{*/},
    commentstyle=\color{gray},
    stringstyle=\color{red},
    morestring=[b]',
    morestring=[b]"
}
\lstdefinelanguage{SVA}{
    keywords={},
    otherkeywords={
    for, begin, end, genvar, generate, logic, property, rose, endproperty, not, assert, posedge, always, if, iff, else
    },
    keywordstyle=\color{blue}\bfseries,
    keywords=[2]{Hierarchy},
    keywordstyle=[2]\color{darkgreen}\bfseries,
    identifierstyle=\color{black},
    sensitive=false,
    comment=[l]{//},
    morecomment=[s]{/*}{*/},
    commentstyle=\color{gray},
    stringstyle=\color{red}
}

\usepackage{tikz}
\usetikzlibrary{patterns,arrows,decorations.pathreplacing}
\usetikzlibrary{arrows.meta}
\usetikzlibrary{positioning}
\usetikzlibrary{positioning,shadows,trees}

\usepackage{tikz-timing}[2009/12/09]
\usetikztiminglibrary{overlays}
\tikztimingmetachar{A}{1.0D{\texttt{ACT}}}
\tikztimingmetachar{P}{1.0D{\texttt{PRE}}}
\tikztimingmetachar{X}{1.0D{\texttt{NOP}}}
\tikztimingmetachar{R}{1.0D{\texttt{RD}}}
\tikztimingmetachar{W}{1.0D{\texttt{WR}}}
\tikztimingmetachar{O}{1.0D{\texttt{NOP}}}

\usepackage{multirow}

\begin{document}

\title{Unveiling the Real Performance of LPDDR5 Memories}

\author{Lukas Steiner}
\orcid{0000-0003-2677-6475}
\affiliation{%
  \institution{Microelectronic Systems Design Research Group, TU Kaiserslautern}
  \streetaddress{Erwin-Schrödinger-Straße 12}
  \city{Kaiserslautern}
  \country{Germany}
  \postcode{67663}
}
\email{lsteiner@eit.uni-kl.de}

\author{Matthias Jung}
\orcid{0000-0003-0036-2143}
\affiliation{%
  \institution{Fraunhofer Institute for Experimental Software Engineering (IESE)}
  \streetaddress{Fraunhofer-Platz 1}
  \city{Kaiserslautern}
  \country{Germany}
  \postcode{67663}
}
\email{matthias.jung@iese.fraunhofer.de}

\author{Michael Huonker}
\affiliation{%
  \institution{Design High-Computing Platforms, Mercedes-Benz}
  \streetaddress{}
  \city{Sindelfingen}
  \country{Germany}
  \postcode{1}
}
\email{michael.huonker@mercedes-benz.com}

\author{Norbert Wehn}
\orcid{0000-0002-9010-086X}
\affiliation{%
  \institution{Microelectronic Systems Design Research Group, TU Kaiserslautern}
  \streetaddress{Erwin-Schrödinger-Straße 12}
  \city{Kaiserslautern}
  \country{Germany}
  \postcode{67663}
}
\email{wehn@eit.uni-kl.de}

\renewcommand{\shortauthors}{Steiner et al.}

\begin{abstract}
LPDDR5 is the latest low-power DRAM standard and expected to be used in various application fields. 
The vendors have published promising peak bandwidths up to 50\,\% higher than those of the predecessor LPDDR4.
In this paper we evaluate the best-case and worst-case real bandwidth utilization of different LPDDR5 configurations and compare the results to corresponding LPDDR4 configurations. 
We also show that an upgrade from LPDDR4 to LPDDR5 does not always bring a bandwidth advantage and that some LPDDR5 configurations should be avoided for specific workloads.
\end{abstract}

%


\maketitle
\section{Introduction}
The latest low-power DRAM standard LPDDR5~\cite{box_19} was released by JEDEC in 2019. 
In the meanwhile, LPDDR5 has gradually replaced its predecessor LPDDR4~\cite{box_14} and is used in various markets including mobile, automotive, AI and 5G~\cite{elsnik_20}. 
However, since the standard comes with lots of changes including a completely new clocking scheme, the memory controllers also have to be largely redesigned. 
Thus, many companies are faced with the decision of whether to upgrade their existing system designs or continue with the well-established predecessor for their target applications. 
However, taking such a decision on the basis of performance numbers by DRAM vendors is difficult as they usually only publish peak bandwidths~\cite{halee_20,leechi_21,samsung1,micron1,hynix1}.
The real bandwidth utilization can vary significantly for different workloads and DRAM device configurations~\cite{gholi_19,stejun_21}.

Therefore, we conduct a simulation-based performance evaluation of different LPDDR5 configurations for sequential and random accesses, which are the best-case and worst-case input patterns with respect to bandwidth utilization. 
In addition, we vary the ratio of read and write requests and compare the results to the performance of corresponding LPDDR4 configurations. 
Finally, we derive several key observations to aid the selection of a suitable memory subsystem. 
All our experiments are carried out with the cycle-accurate DRAM simulation framework DRAMSys\cite{stejun_20}.

The remaining paper is structured as follows: Section~\ref{sec:lpddr5} gives an overview of the LPDDR5 standard and its innovations compared to the predecessor. Section~\ref{sec:evaluation} describes the setup used for evaluation before the simulation results are presented in Section~\ref{sec:results}. Finally, Section~\ref{sec:conclusion} concludes the paper.








\section{LPDDR5 Overview}\label{sec:lpddr5}
%

As with every new JEDEC standard release also LPDDR5 enhances the most important DRAM key parameters, i.e., bandwidth, storage capacity and power consumption. 
LPDDR5 devices can operate at data rates up to 6400\,MT/s\footnote{The latest revision of LPDDR5 (called LPDDR5X) even specifies data rates up to 8533\,MT/s.} compared to 4266\,MT/s for LPDDR4. 
Since the clock signal would contribute a considerable amount to the total power at such high data rates, LPDDR5 devices use a slower continuously running command clock and a faster data clock that is disabled during idle times.
Depending on the data rate the ratio between both clock signals can either be set to 2:1 (up to 3200\,MT/s) or 4:1.
To compensate for the slow command clock, commands are now transmitted at double data rate.
While an LPDDR4 channel is always comprised of 8 banks, LPDDR5 introduces three different bank modes, which can be selected at power-up. For data rates up to 3200\,MT/s a channel is comprised of 16 banks by default (16B mode). 
For higher data rates these 16 banks are divided into 4 bank groups (called BG mode). 
Alternatively, a mode with 8 banks (8B mode) can be chosen where two internal banks are merged to form a single bank with double the capacity visible on the interface. 
This mode is supported at all data rates.  
The prefetch for 16B and BG mode is 16n with a default burst length of 16 and an optional burst length of 32.
With 8B mode enabled the prefetch is increased to 32n and the burst length is also fixed to 32.
One constraint of the BG mode in combination with burst length 32 is that the data will be transmitted over the bus in an interleaved fashion, i.e., it is divided into two halves with a gap in between.

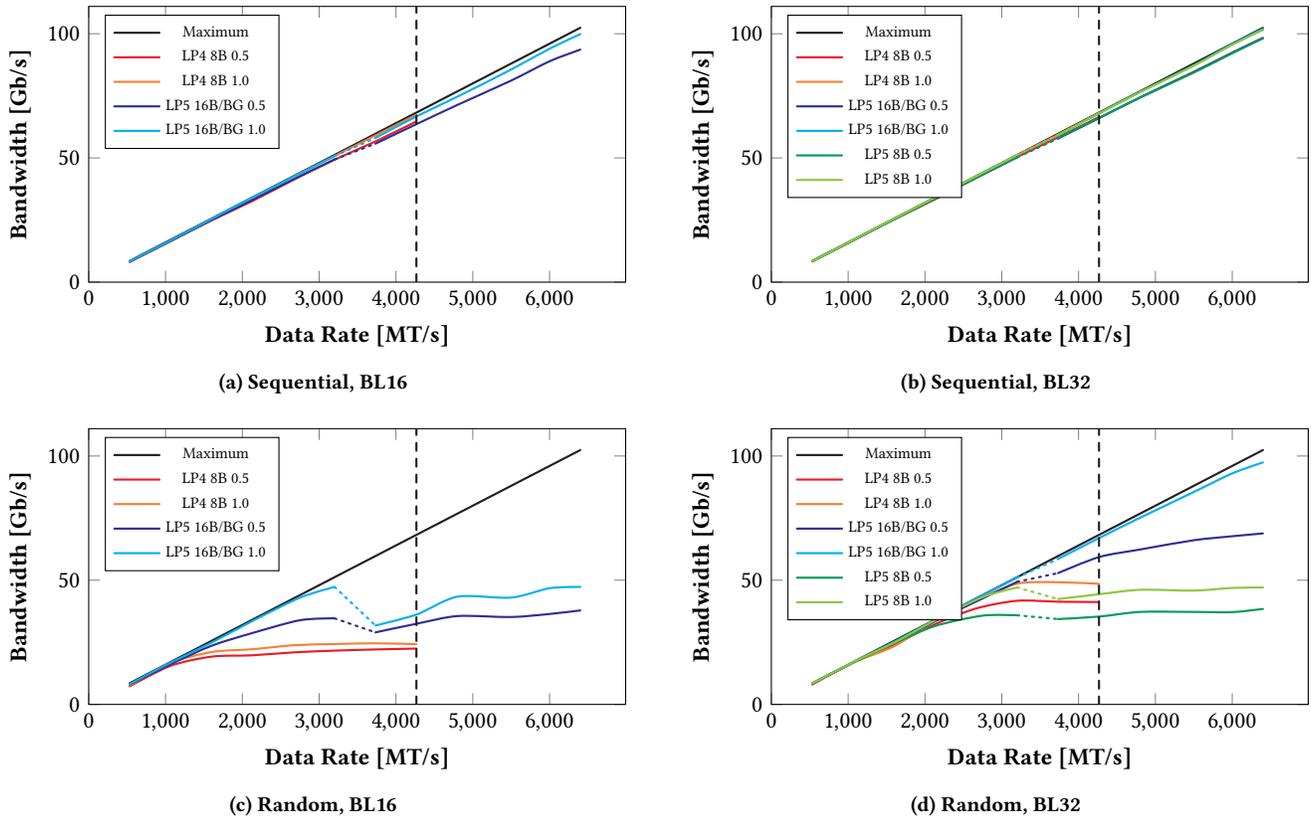
\begin{figure*}[h!]
    \centering
    \begin{subfigure}[b]{0.49\textwidth}
        \centering
        \begin{tikzpicture}
        \begin{axis}[
        	ylabel={\textbf{Bandwidth [Gb/s]}},
        	xlabel={\textbf{Data Rate [MT/s]}},
            grid=minor,
            width = \textwidth,
            height = 5.25cm,
            xmin = 0,
            xmax = 6990,
            ymin = 0,
            ymax = 111,
            legend style={legend pos=north west, font=\scriptsize}
        ]
        
        \addplot[Black, thick, line cap=round, smooth] coordinates {
            (0533,8.528)
            (1066,17.056)
            (1600,25.6)
            (2133,34.128)
            (2666,42.656)
            (3200,51.2)
            (3733,59.728)
            (4266,68.256)
            (4800,76.8)
            (5500,88)
            (6000,96)
            (6400,102.4)
        };
        
        \addplot[Red, thick, line cap=round, smooth] coordinates {
           (533,8.1)
           (1066,16.49)
           (1600,24.88)
           (2133,32.69)
           (2666,41.22)
           (3200,49.3)
           (3733,56.64)
           (4266,64.7)
 
        };
        
        \addplot[Orange, thick, line cap=round, smooth] coordinates {
            ( 533,8.26)
            (1066,16.67)
            (1600,25.44)
            (2133,33.68)
            (2666,42.45)
            (3200,50.77)
            (3733,59.04)
            (4266,67.39)
        };
        
        \addplot[Blue, thick, line cap=round, smooth] coordinates {
            (0533,8.25 )
            (1067,16.6 )
            (1600,24.99)
            (2133,33.12)
            (2750,42.64)
            (3200,49.39)
        };

        \addplot[ProcessBlue, thick, line cap=round, smooth] coordinates {
            (0533,8.38)
            (1067,16.82)
            (1600,25.43)
            (2133,34.09)
            (2750,43.65)
            (3200,50.77)
        };

        \addplot[Blue, thick, line cap=round, smooth] coordinates {
            (3733,55.74)
            (4267,63.6 )
            (4800,71.35)
            (5500,81.15)
            (6000,88.92)
            (6400,93.61)
        };

        \addplot[ProcessBlue, thick, line cap=round, smooth] coordinates {
            (3733,58.07)
            (4267,66.74)
            (4800,74.66)
            (5500,85.61)
            (6000,93.98)
            (6400,99.86)
        };
        
        \addplot[Blue, thick, line cap=round, smooth, dotted] coordinates {
            (3200,49.39)
            (3733,55.74)
        };

        \addplot[ProcessBlue, thick, line cap=round, smooth, dotted] coordinates {
            (3200,50.77)
            (3733,58.07)
        };
        
        \addplot[Black, thick, line cap=round, smooth, dashed] coordinates {
            (4266, 0)
            (4266, 111)
        };
        
        \legend{
            Maximum,%
            LP4 8B 0.5,%
            LP4 8B 1.0,%
            LP5 16B/BG 0.5,%
            LP5 16B/BG 1.0,%
        }
        \end{axis}
        \end{tikzpicture}
        \caption{Sequential, BL16}
        \label{fig:bw:seq_bl16}
    \end{subfigure}
    \hfill
    \begin{subfigure}[b]{0.49\textwidth}
        \centering
        \begin{tikzpicture}
        \begin{axis}[
        	ylabel={\textbf{Bandwidth [Gb/s]}},
        	xlabel={\textbf{Data Rate [MT/s]}},
            grid=minor,
            width = \textwidth,
            height = 5.25cm,
            xmin = 0,
            xmax = 6990,
            ymin = 0,
            ymax = 111,
            legend style={legend pos=north west, font=\scriptsize}
        ]
        
        \addplot[Black, thick, line cap=round, smooth] coordinates {
            (0533,8.528)
            (1066,17.056)
            (1600,25.6)
            (2133,34.128)
            (2666,42.656)
            (3200,51.2)
            (3733,59.728)
            (4266,68.256)
            (4800,76.8)
            (5500,88)
            (6000,96)
            (6400,102.4)
        };
        
        \addplot[Red, thick, line cap=round, smooth] coordinates {
            (533 ,8.31 )
            (1066,16.79)
            (1600,25.31)
            (2133,33.39)
            (2666,42   )
            (3200,50.36)
            (3733,58.32)
            (4266,66.48)
        };
        
        \addplot[Orange, thick, line cap=round, smooth] coordinates {
            (533 ,8.38)
            (1066,16.8)
            (1600,25.53)
            (2133,34.08)
            (2666,42.55)
            (3200,51.12)
            (3733,59.32)
            (4266,68.06)
        };

        \addplot[Blue, thick, line cap=round, smooth] coordinates {
            (0533,8.35)
            (1067,16.91)
            (1600,25.33)
            (2133,33.66)
            (2750,43.28)
            (3200,50.3)
        };

        \addplot[ProcessBlue, thick, line cap=round, smooth] coordinates {
            (0533,8.38  )
            (1067,17.03 )
            (1600,25.52 )
            (2133,34.07 )
            (2750,43.92 )
            (3200,51.09 )
        };

        \addplot[ForestGreen, thick, line cap=round, smooth] coordinates {
            (0533,8.36)
            (1067,16.93)
            (1600,25.34)
            (2133,33.64)
            (2750,43.37)
            (3200,50.32)
        };

        \addplot[LimeGreen, thick, line cap=round, smooth] coordinates {
            (0533,8.38  )
            (1067,17.03 )
            (1600,25.55 )
            (2133,34.07 )
            (2750,43.93 )
            (3200,51.12 )
        };

        \addplot[Blue, thick, line cap=round, smooth] coordinates {
            (3733,57.84)
            (4267,66.02)
            (4800,74.38)
            (5500,84.65)
            (6000,92.4)
            (6400,98.34)
        };

        \addplot[ProcessBlue, thick, line cap=round, smooth] coordinates {
            (3733,59.23 )
            (4267,67.95 )
            (4800,76.48 )
            (5500,87.32 )
            (6000,95.81 )
            (6400,101.97)
        };

        \addplot[ForestGreen, thick, line cap=round, smooth] coordinates {
            (3733,58.1)
            (4267,66.14)
            (4800,74.27)
            (5500,84.35)
            (6000,92.12)
            (6400,98.1)
        };

        \addplot[LimeGreen, thick, line cap=round, smooth] coordinates {
            (3733,59.3  )
            (4267,67.97 )
            (4800,76.53 )
            (5500,87.1  )
            (6000,95.53 )
            (6400,101.52)
        };
        
        \addplot[Blue, thick, line cap=round, smooth, dotted] coordinates {
            (3200,50.3)
            (3733,57.84)
        };

        \addplot[ProcessBlue, thick, line cap=round, smooth, dotted] coordinates {
            (3200,51.09 )
            (3733,59.23 )
        };

        \addplot[ForestGreen, thick, line cap=round, smooth, dotted] coordinates {
            (3200,50.32)
            (3733,58.1)
        };

        \addplot[LimeGreen, thick, line cap=round, smooth, dotted] coordinates {
            (3200,51.12 )
            (3733,59.3  )
        };
        
        \addplot[Black, thick, line cap=round, smooth, dashed] coordinates {
            (4266, 0)
            (4266, 111)
        };
        
        \legend{
            Maximum,%
            LP4 8B 0.5,%
            LP4 8B 1.0,%
            LP5 16B/BG 0.5,%
            LP5 16B/BG 1.0,%
            LP5 8B 0.5,
            LP5 8B 1.0,
        }
        \end{axis}
        \end{tikzpicture}
        \caption{Sequential, BL32}
        \label{fig:bw:seq_bl32}
    \end{subfigure}
    \vskip\baselineskip
    \begin{subfigure}[b]{0.49\textwidth}
        \centering
        \begin{tikzpicture}
        \begin{axis}[
        	ylabel={\textbf{Bandwidth [Gb/s]}},
        	xlabel={\textbf{Data Rate [MT/s]}},
            grid=minor,
            width = \textwidth,
            height = 5.25cm,
            xmin = 0,
            xmax = 6990,
            ymin = 0,
            ymax = 111,
            legend style={legend pos=north west, font=\scriptsize}
        ]
        
        \addplot[Black, thick, line cap=round, smooth] coordinates {
            (0533,8.528)
            (1066,17.056)
            (1600,25.6)
            (2133,34.128)
            (2666,42.656)
            (3200,51.2)
            (3733,59.728)
            (4266,68.256)
            (4800,76.8)
            (5500,88)
            (6000,96)
            (6400,102.4)
        };
        
        \addplot[Red, thick, line cap=round, smooth] coordinates {
            (533 ,7.28 )
            (1066,15.55)
            (1600,19.23)
            (2133,19.78)
            (2666,20.95)
            (3200,21.65)
            (3733,22.12)
            (4266,22.48)
        };
        
        \addplot[Orange, thick, line cap=round, smooth] coordinates {
            (533 ,7.62)
            (1066,16.71)
            (1600,21.09)
            (2133,22.22)
            (2666,23.83)
            (3200,24.3)
            (3733,24.6)
            (4266,24.27)
        };
        
        \addplot[Blue, thick, line cap=round, smooth] coordinates {
            (0533,8.04)
            (1067,16.18)
            (1600,23.62)
            (2133,28.76)
            (2750,33.9)
            (3200,34.65)
        };

        \addplot[ProcessBlue, thick, line cap=round, smooth] coordinates {
            (0533,8.18 )
            (1067,16.84)
            (1600,24.87)
            (2133,33.61)
            (2750,42.97)
            (3200,47.31)
        };

        \addplot[Blue, thick, line cap=round, smooth] coordinates {
            (3733,29)
            (4267,32.47)
            (4800,35.57)
            (5500,35.18)
            (6000,36.41)
            (6400,37.83)
        };

        \addplot[ProcessBlue, thick, line cap=round, smooth] coordinates {
            (3733,31.77)
            (4267,36.17)
            (4800,43.37)
            (5500,43   )
            (6000,46.78)
            (6400,47.31)
        };
        
        \addplot[Blue, thick, line cap=round, smooth, dotted] coordinates {
            (3200,34.65)
            (3733,29)
        };

        \addplot[ProcessBlue, thick, line cap=round, smooth, dotted] coordinates {
            (3200,47.31)
            (3733,31.77)
        };
        
        \addplot[Black, thick, line cap=round, smooth, dashed] coordinates {
            (4266, 0)
            (4266, 111)
        };
        
        \legend{
            Maximum,%
            LP4 8B 0.5,%
            LP4 8B 1.0,%
            LP5 16B/BG 0.5,%
            LP5 16B/BG 1.0,%
        }
        \end{axis}
        \end{tikzpicture}
        \caption{Random, BL16}
        \label{fig:bw:rdm_bl16}
    \end{subfigure}
    \hfill
    \begin{subfigure}[b]{0.49\textwidth}
        \centering
        \begin{tikzpicture}
        \begin{axis}[
        	ylabel={\textbf{Bandwidth [Gb/s]}},
        	xlabel={\textbf{Data Rate [MT/s]}},
            grid=minor,
            width = \textwidth,
            height = 5.25cm,
            xmin = 0,
            xmax = 6990,
            ymin = 0,
            ymax = 111,
            legend style={legend pos=north west, font=\scriptsize}
        ]
        
        \addplot[Black, thick, line cap=round, smooth] coordinates {
            (0533,8.528)
            (1066,17.056)
            (1600,25.6)
            (2133,34.128)
            (2666,42.656)
            (3200,51.2)
            (3733,59.728)
            (4266,68.256)
            (4800,76.8)
            (5500,88)
            (6000,96)
            (6400,102.4)
        };
        
        \addplot[Red, thick, line cap=round, smooth] coordinates {
            (533 ,8.01 )
            (1066,16.84)
            (1600,24.11)
            (2133,32.32)
            (2666,38.63)
            (3200,41.72)
            (3733,41.35)
            (4266,41.21)
        };
        
        \addplot[Orange, thick, line cap=round, smooth] coordinates {
            (533 ,8.41)
            (1066,16.88)
            (1600,23.44)
            (2133,33.67)
            (2666,41.89)
            (3200,48.58)
            (3733,49.19)
            (4266,48.54)
        };

        \addplot[Blue, thick, line cap=round, smooth] coordinates {
            (0533,8.36)
            (1067,16.91)
            (1600,24.95)
            (2133,33.5)
            (2750,42.65)
            (3200,49.36)
        };

        \addplot[ProcessBlue, thick, line cap=round, smooth] coordinates {
            (0533,8.41 )
            (1067,17.03)
            (1600,25.19)
            (2133,34.01)
            (2750,43.88)
            (3200,51.06)
        };

        \addplot[ForestGreen, thick, line cap=round, smooth] coordinates {
            (0533,8.4)
            (1067,16.89)
            (1600,24.61)
            (2133,31.58)
            (2750,35.74)
            (3200,35.88)
        };

        \addplot[LimeGreen, thick, line cap=round, smooth] coordinates {
            (0533,8.44 )
            (1067,17.01)
            (1600,24.91)
            (2133,33.88)
            (2750,43.17)
            (3200,47.07)
        };

        \addplot[Blue, thick, line cap=round, smooth] coordinates {
            (3733,52.86)
            (4267,59.33)
            (4800,62.36)
            (5500,66.09)
            (6000,67.7)
            (6400,68.83)
        };

        \addplot[ProcessBlue, thick, line cap=round, smooth] coordinates {
            (3733,58.56)
            (4267,66.98)
            (4800,75.14)
            (5500,85.51)
            (6000,93   )
            (6400,97.42)
        };

        \addplot[ForestGreen, thick, line cap=round, smooth] coordinates {
            (3733,34.36)
            (4267,35.36)
            (4800,37.24)
            (5500,37.23)
            (6000,37.15)
            (6400,38.43)
        };

        \addplot[LimeGreen, thick, line cap=round, smooth] coordinates {
            (3733,42.41)
            (4267,44.36)
            (4800,46.14)
            (5500,45.82)
            (6000,46.88)
            (6400,47.07)
        };
        
        \addplot[Blue, thick, line cap=round, smooth, dotted] coordinates {
            (3200,49.36)
            (3733,52.86)
        };

        \addplot[ProcessBlue, thick, line cap=round, smooth, dotted] coordinates {
            (3200,51.06)
            (3733,58.56)
        };

        \addplot[ForestGreen, thick, line cap=round, smooth, dotted] coordinates {
            (3200,35.88)
            (3733,34.36)
        };

        \addplot[LimeGreen, thick, line cap=round, smooth, dotted] coordinates {
            (3200,47.07)
            (3733,42.41)
        };
        
        \addplot[Black, thick, line cap=round, smooth, dashed] coordinates {
            (4266, 0)
            (4266, 111)
        };
        
        \legend{
            Maximum,%
            LP4 8B 0.5,%
            LP4 8B 1.0,%
            LP5 16B/BG 0.5,%
            LP5 16B/BG 1.0,%
            LP5 8B 0.5,
            LP5 8B 1.0,
        }
        \end{axis}
        \end{tikzpicture}
        \caption{Random, BL32}
        \label{fig:bw:rdm_bl32}
    \end{subfigure}
    \caption{Real Bandwidth Utilization of LPDDR4/LPDDR5 Configurations for Different Input Traffics}
    \label{fig:bw}
\end{figure*}

\section{Evaluation}\label{sec:evaluation}
For an extensive performance evaluation all possible data rates, bank modes (16B/BG/8B) and burst lengths (BL16/BL32) are taken into consideration. 
The simulated LPDDR5 devices have a density of 16\,Gb and a single channel with a width of 16\,bits, while the density of the LPDDR4 devices is only 8\,Gb but with an identical channel configuration.
For LPDDR5 speed grades up to 3200\,MT/s the 2:1 clocking ratio is selected, for higher speed grades the ratio is changed to 4:1. 
All devices are operated with per-bank refresh. 
The controller uses a first-ready, first-come-first-serve (FR-FCFS) scheduler~\cite{rixdal_00}, 64-entry read and write queues and a row-column-bank\mbox{(-bank group)} address mapping. 
As input stimuli either sequential or random traffic is considered. 
In addition, the ratio between read and write requests is either set to 0.5 (same number of reads and writes) or 1.0 (only reads). 
For sequential traffic the open-page policy is chosen, for random traffic the closed-page policy~\cite{hanaga_14}.

\section{Results}\label{sec:results}
For sequential input traffic (see Figures~\ref{fig:bw:seq_bl16} and \ref{fig:bw:seq_bl32}) we observe that irrespective of the burst length and bank mode both LPDDR4 and LPDDR5 devices achieve bandwidths very close to the theoretical maximum. Only for the highest data rates and traffic with mixed reads and writes the bandwidth slightly drops up to a maximum loss of around 8.5\,\% (LPDDR5 with BL16), which is still a very reasonable result for a single-rank memory channel. The main reasons for the low performance drop are the large number of row hits and the use of per-bank refresh, which allows refresh operations to be executed in the background with blocking only one or two banks and not the whole device at once. Thus, the results also show that for target data rates below 4266\,MT/s an upgrade from LPDDR4 to LPDDR5 does not bring any advantage at all from a bandwidth perspective.

For random input traffic, however, the experiments reveal an entirely different behavior. 
With a burst length of 16 LPDDR4 cannot keep up the bandwidth utilization at all and already starts saturation at data rates of 1600\,MT/s, which is only around one third of the maximum (see Figure~\ref{fig:bw:rdm_bl16}). 
At the highest data rate 4266\,MT/s the device only achieves 33\,\% of the maximum bandwidth for mixed traffic and 35,5\,\% for pure read traffic. 
The reason for the poor results is the low bank parallelism of LPDDR4. 
LPDDR5, on the other hand, comes with 16 banks and therefore keeps the bandwidth utilization up also at higher data rates. 
However, at the transition from the 2:1 to 4:1 clock ratio (dotted line between 3200\,MT/s and 3733\,MT/s) the bandwidth drops significantly and only reaches the previous peak bandwidth again at data rates of 6400\,MT/s. 
This behavior has two main reasons. 
First, with 4:1 clocking the command bus operates at very low frequencies compared to the data bus and a burst of length 16 only occupies the data bus for two cycles in relation to the command clock. 
With random traffic each access results in a row miss, which translates into an activate command and a read/write command with auto-precharge. 
For LPDDR5 these two commands take a total of 3 clock on the command bus. 
Thus, the command bus occupation limits the bandwidth utilization to 66\,\% of the upper limit. 
Second, the low frequency leads to strongly overestimated timings because analog values have to be rounded to a multiple of one command clock cycle.

When doubling the burst length both LPDDR4 and LPDDR5 perform significantly better, since the number of row misses per time unit is halved. LPDDR4 and LPDDR5 in 8B mode show a similar behavior (saturation at 3200\,MT/s) with a slight advantage for LPDDR4 because of its shorter timings (especially write recovery time) and less rounding. LPDDR5 in 16B/BG mode keeps the bandwidth close to the maximum even up to the highest data rate for pure read traffic and still reaches 67\,\% of the maximum bandwidth for mixed traffic due to the higher bank parallelism. 

The results can be summarized in the following key observations:
\begin{itemize}
    \item For sequential traffic an upgrade from LPDDR4 to LPDDR5 only brings an advantage from a bandwidth perspective if data rates above 4266\,MT/s are targeted.
    \item For random traffic and a burst length of 16 the 4:1 clock ratio of LPDDR5 should be avoided.
    \item For random traffic LPDDR5 in 16B/BG mode achieves much higher bandwidths than LPDDR4.
    \item For random traffic the burst length should be selected as high as possible.
\end{itemize}

\section{Conclusion}\label{sec:conclusion}
In this paper we have analyzed the real bandwidth utilization of different LPDDR5 configurations and compared the results to the predecessor LPDDR4. We found out that for workloads with sequential traffic upgrading from LPDDR4 to LPDDR5 is not always worth it, while for random traffic LPDDR5 outperforms its predecessor in most cases. For future work we plan to extend the evaluation to a full-system simulation setup and analyze the performance of LPDDR5 for real application benchmarks.

\begin{acks}
This work was supported within the Fraunhofer and DFG cooperation programme (Grant no. 248750294) and supported by the Fraunhofer High Performance Center for Simulation- and Software-based Innovation.
\end{acks}

\bibliographystyle{ACM-Reference-Format}


%
%
%
%
%
%
%
%

\end{document}